# رؤوس الموضوعات المقننة في بيئة الفهارس الآلية على الخط المباشر: دراسة تطبيقية على عينة من التسجيلات العربية

# Authorized Subject Headings in the Online Automatic catalog Environment: An Empirical Study on a Sample of Arabic Records

الملخص العربي:

لرؤوس الموضوعات المقننة أهمية كبيرة بالفهارس الآلية نظرًا لما تمثله من أهمية للبحث الموضوعي. وتهدف هذه الدراسة إلى قياس جودة مجموعة من رؤوس الموضوعات المقننة بعينة من التسجيلات الببليوجرافية العربية على فهرس مكتبات الجامعات المصرية من خلال التعرف على أهم الممارسات والسياسات والاجراءات المتبعة، والأدوات المستخدمة. بالإضافة إلى تقييم الإمكانات الفعلية للقوائم، والمكانز، والأدلة الإرشادية التي تم الاستناد إليها في وضع نقاط الإتاحة الموضوعية. استخدمت الدراسة كلا المنهجين الوصفي التحليلي والمنهج التقييمي لتحقيق أهداف الدراسة.

الملخص الإنجليزي:

subject headings are very important to machine catalogs, given the importance of thematic research. This study aims to measure the quality of a group of authorized subject headings with a sample of Arabic bibliographic records on the catalog of Egyptian university libraries by identifying the most important practices, policies, procedures followed, and tools used. In addition to assessing the actual capabilities of lists, thesaurus, and guidelines that were used in establishing thematic availability points. The study used both the descriptive analytical and evaluation approaches to achieve the study objectives.

الكلمات المفتاحية

رؤوس الموضوعات – نقاط الإتاحة – الفهارس الآلية – الفهارس الموحدة – التسجيلات الببليوجرافية.

# رؤوس الموضوعات المقننة في بيئة الفهارس الآلية على الخط المباشر: دراسة تطبيقية على عينة من التسجيلات العربية

# Authorized Subject Headings in the Online Automatic catalog Environment: An Empirical Study on a Sample of Arabic Records

أحمد عمار حسين همام

المحاضر بجامعة الحدود الشمالية

Ahmed.Hassein@nbu.edu.sa

Ahmednbu2@gmail.com

## 0/1 تمهيد:

مع بداية التسعينيات من القرن العشرين توافرت في العالم العربي مجموعة من البرمجيات العربية والمعربة، التي تستطيع أن تساعد في أداء معظم أو كل الوظائف والخدمات، التي تقوم بها المكتبات ومراكز المعلومات فكان ذلك عاملًا حيويًا، لدفع حركة ميكنة المكتبات ومراكز المعلومات العربية خطوات واسعة إلى الأمام، ونتج عن تسخير هذه التكنولوجيا في المكتبات ومراكز المعلومات ما يعرف بالفهارس الموحدة للمكتبات، ومن أمثلتها الفهرس العربي الموحد، والفهرس الموحد لاتحاد المكتبات الجامعية المصرية الذي تم إنشاؤه تحت مظلة مشروع ميكنة مكتبات الجامعات المصرية.

في هذا الإطار وخاصة في ظل وجود مجموعة من المكتبات المختلفة والتي لابد أن تتشارك في تسجيلة واحدة كان لابد من وجود نوع من توحيد الممارسة، وذلك بهدف توحيد المنتج النهائي المتمثل في التسجيلة الببليوجرافية؛ ونتيجة لذلك ظهرت مجموعة من السياسات والإجراءات والممارسات من قبل المكتبيين فيما يتعلق بالتحليل الموضوعي، وتوفير نقاط الإتاحة الموضوعية المقننة، من خلال اختيار وصياغة رؤوس الموضوعات المناسبة والتي تعبر تعبيرًا دقيقًا عن المحتوى الموضوعي لأوعية المعلومات، مما يسهل عملية وصول المستفيد إلى ما يحتاج إليه من معلومات.

تشير العديد من الدراسات التي أجريت حول سلوك المستفيدين في البحث عن المعلومات أن أغلب المستفيدين يلجأون إلى البحث عن أوعية معلومات غير معروفة، عبر البحث باستخدام الموضوع، مقارنة بعدد أقل من المستفيدين، الذين يبحثون عن أوعية معلومات معروفة عبر

المداخل الأخرى المتاحة للبحث. وعلى الرغم من أهمية الوصول إلى مواد المعلومات بالموضوع، إلا أن توفير نقاط إتاحة موضوعية مقننة (رؤوس موضوعات) لمواد المعلومات ليس بالأمر الهين، بل هو من أصعب المهام التي تمر بها العمليات الفنية في المكتبات، ففي ظل الثورة المعلوماتية الحالية أصبحت الموضوعات تتسم بالتشابك والتعقيد والتداخل مما صعّب مهمة التحديد الدقيق للموضوعات، إضافة إلى صعوبة وضع قواعد حاكمة وصارمة لاختيار وصياغة نقاط الإتاحة الموضوعية المقننة، مما انعكس سلبًا على أداء الفهارس من الناحية الموضوعية وخاصة في فهارس مكتباتنا العربية.

ترتبط جودة نقاط الإتاحة الموضوعية المقننة بفهارس مكتباتنا العربية بعدة عوامل؛ أبرزها صعوبة الوصول إلى المعايير العالمية المتعارف عليها لمعدلات وأعداد رؤوس الموضوعات المستخدمة في الفهارس ومدى تطبيق القواعد المقننة لعملية اختيار وصياغة نقاط الإتاحة الموضوعية بشكل عام، وفي بيئة الفهارس الآلية بشكل خاص.

<u>0/2 أهمية الموضوع ومبررات اختياره :</u>

ترجع أهمية الدراسة الحالية ومضمونها للأسباب التالية:

- تنبع أهمية هذه الدراسة من أهمية الفهرس الموحد لاتحاد المكتبات الجامعية المصرية واعتباره مشروع قومي يخدم الباحثين والمتخصصين وخاصة بالجامعات والمؤسسات العلمية في مصر، والذين ينصب جهدهم في البحث بالموضوع عبر منافذ البحث المتاحة عبر الفهرس الموحد لاتحاد المكتبات الجامعية المصرية.

- أهمية التحليل الموضوعي في ظل الفهارس الآلية الموحدة، وتشارك المكتبات المساهمة في تسجيلة واحدة؛ مما يستدعي توحيد الممارسات والسياسات والإجراءات المتبعة لاختيار وصياغة رؤوس الموضوعات، فضلًا عن مهارات التعامل مع النظم الآلية في المكتبات فيما يتعلق بالعمليات الفنية وخاصة التحليل الموضوعي.

- أهمية نقاط الإتاحة الموضوعية المقننة وتأثير جودتها على كفاءة الاسترجاع الموضوعي في ظل الفهارس الآلية الموحدة.

- تعد تلبية احتياجات المستفيد، وإمداده بكل ما يحتاج إليه من مواد المعلومات التي يحتاج إليها، من الأهداف الحيوية للمكتبات، وخاصة تلك التي تتشارك مع بعضها البعض في فهرس واحد، فعلى الرغم من توفير الفهارس الآلية للعديد من نقاط الإتاحة التي يتم البحث من خلالها للوصول إلى مواد المعلومات إلا أن توفير نقاط الإتاحة الموضوعية، يكتسب أهمية خاصة، من حيث كونه أفضل السبل التي يلجأ إليها المستفيد للحصول على مواد المعلومات، وهذا ما أكدت عليه العديد من الدراسات.

3/0 مشكلة الدراسة :

برزت مشكلةُ الدراسةِ، واتضحت أهميتَها من خلال عمل الباحث بمشروع ميكنة المكتبات الجامعية المصرية كاستشاري لضبط الجودة حيث تمركزت عناصرَ المشكلة فيما تم ملاحظتُهُ من القصورِ الواضحِ في عملية اختيار وصياغةِ رؤوسِ الموضوعات. ووجودُ قصورٍ في السياسات وأخطاء في الممارسات فيما يتعلقُ بالتحليل الموضوعي. بالإضافة إلى ضعفٍ وقصورٍ وعدم فعالية وكفاية أدوات العمل الفنية المستخدمة للتحليل الموضوعي مثل المكانز ورؤوس الموضوعات والأدلة الإرشادية. فضلًا عن غياب ملف استناد آلي للموضوعات حيث يقتصر الأمر على القائمة الاستنادية التي يوفرها النظام الآلي بالفهرس دون تفعيل النظام الفرعي للضبط الاستنادي.

كما برزت مشكلة الدراسة مما تم ملاحظته من عدم تحري الدقة من قِبَل المفهرسين الموضوعيين في تعيين رؤوس الموضوعات بحيث تعكس المحتوى الموضوعي لهذا العنصر. إما لضيق أوقاتهم أو لعدم إلمامهم بالمجال الموضوعي للعمل المفهرس، أو عدم مراجعة المعلومات ذات الصلة مثل المقدمة والتمهيد أو جدول المحتويات. بالإضافة إلى ما يحدث في كثير من الأحيان من تعيين رؤوس موضوعات واسعة أو غير دقيقة.

استنادًا إلى كل ما سبق وإيمانًا من الباحث بأهمية الاسترجاع الموضوعي بدت الحاجة الماسة إلى دراسة وتقييم جودة نقاط الإتاحة الموضوعية المقننة في ظل الفهارس الآلية الموحدة بالتطبيق على الفهرس الآلي الموحد لاتحاد المكتبات الجامعية المصرية.

4/0 أهداف الدراسة :

تهدف الدراسة إلى قياس جودة نقاط الإتاحة الموضوعية المقننة للتسجيلات الببليوجرافية العربية بالفهرس الموحد لمكتبات الجامعات المصرية من خلال ما يلي:

- التعرف على أهم الممارسات والسياسات والإجراءات المتبعة، والأدوات المستخدمة بمشروع الفهرس الموحد لاتحاد المكتبات الجامعية المصرية؛ لتوفير نقاط الإتاحة الموضوعية المقننة بالتسجيلات الببليوجرافية سواء بوحدة المكتبة الرقمية، أو بالمكتبات المساهمة في مشروع الفهرس.

- دراسة وتقييم الإمكانات الفعلية للقوائم، والمكانز، والأدلة الإرشادية التي تم الاستناد إليها في وضع نقاط الإتاحة الموضوعية المقننة بالتسجيلات الببليوجرافية؛ لتحديد درجة الجودة في هذه القوائم والمكانز والأدلة، ومدى كفايتها للتحليل الموضوعي لأوعية المعلومات بمكتبات الاتحاد.

- دراسة وتقييم الواقع الحالي لبيئة العمل بمشروع الفهرس الموحد لاتحاد المكتبات الجامعية المصرية، ومدى ملائمته لتوفير قوائم الاستناد الموضوعي الخاصة بالفهرس؛ لتحديد درجة

الجودة في إعدادها وتنفيذها.

- دراسة السياسات التي تضعها إدارة الفهرس، فيما يتعلق باختيار نقاط الإتاحة الموضوعية؛ لتحديد درجة الجودة في تنفيذ العاملين لهذه السياسات.
- قياس مدى الدقة في اختيار وصياغة رؤوس الموضوعات بالتسجيلات الببليوجرافية للكتب العربية بالفهرس الموحد لمكتبات الجامعات المصرية، ومضاهاتها بأدوات التحليل الموضوعي المعتمدة من قبل وحدة المكتبة الرقمية بالمجلس الأعلى للجامعات.
- قياس مدي الدقة في اختيار وصياغة رؤوس الموضوعات لتوفير نقاط الإتاحة الموضوعية المقننة وفق المعايير والسياسات التي أوصت بها وحدة المكتبة الرقمية بالمجلس الأعلى للجامعات.

### 0/5 تساؤلات الدراسة:

تسعى الدراسة في سبيل تحقيق الأهداف السابقة إلى الإجابة عن التساؤلات التالية:

1- ما أبرز التجارب العالمية والعربية في مجال الفهارس الموحدة؟
2- ما خصائص نقاط الإتاحة الموضوعية المقننة بالفهارس الآلية وفق شكل الاتصال مارك (MARC)؟
3- ما مدى توافر وتحديد السياسات والإجراءات المتبعة في التحليل الموضوعي لأوعية المعلومات، وأثره على جودة نقاط الإتاحة الموضوعية المقننة بفهرس اتحاد المكتبات الجامعية المصرية؟
4- ما المشكلات والصعوبات التي تواجه القائمين على مشروع فهرس اتحاد المكتبات الجامعية المصرية، والتي تؤثر سلبًا على جودة نقاط الإتاحة الموضوعية بالتسجيلات الببليوجرافية بالفهرس؟
5- إلى أي مدى يؤثر عدم الالتزام بأسس ومبادئ اختيار وصياغة رؤوس الموضوعات على الأداء العام لفهرس اتحاد المكتبات الجامعية المصرية، من ناحية الاسترجاع الموضوعي وما هي المؤشرات القياسية والعددية على ذلك؟
6- ما كفاءة رؤوس الموضوعات بفهرس اتحاد المكتبات الجامعية المصرية، وانعكاس ذلك على كل فرع من فروع المعارف البشرية العشرة، وفق تصنيف ديوي العشري (DDC)؟

### 0/6 حدود الدراسة:

❖ <u>الحدود الموضوعية:</u>

تسعى الدراسة إلى التعرف على جودة نقاط الإتاحة الموضوعية بالتسجيلات الببليوجرافية

للمكتبات الأعضاء بمشروع فهرس اتحاد المكتبات الجامعية المصرية، من خلال التعرف على الممارسات، وأهم السياسات والإجراءات المتبعة، سواء بوحدة المكتبة الرقمية – الجهة المشرفة على مشروع فهرس اتحاد المكتبات الجامعية المصرية – أو بالمكتبات الأعضاء بمشروع الفهرس، وأثر ذلك على جودة نقاط الإتاحة الموضوعية المقننة بالتسجيلات الببليوجرافية.

❖ الحدود المكانية:

يضم مشروع الفهرس الموحد لاتحاد المكتبات الجامعية المصرية في عضويته مكتبات المؤسسات الحكومية والمتمثلة في مكتبات الكليات والأقسام والمكتبات المركزية بالجامعات على مستوى جمهورية مصر العربية، حيث بلغ عدد المؤسسات الحكومية ( الجامعات – الأكاديميات – المعاهد) حوالى (30) مؤسسة حكومية بالإضافة إلى مؤسستين خاصتين وهما: ( المعهد التكنولوجي العالي بالعاشر من رمضان – معهد أكتوبر العالي للهندسة والتكنولوجيا)، بالإضافة إلى ثلاث مؤسسات تندرج تحت فئات أخرى هي (الأكاديمية العربية للعلوم المالية والمصرفية – مؤسسة المرأة والذاكرة – المكتب الثقافي بلندن).

بلغ عدد مكتبات تلك المؤسسات حوالي (418) مكتبة من بينها المكتبات المركزية بالجامعات ومكتبات الكليات والمعاهد والأكاديميات، وحتى المكتبات الخاصة ببعض الأفراد وتم إهدائها للأقسام والكليات بالجامعات.

الجدير بالذكر أن أماكن مكتبات تلك المؤسسات بمختلف أنواعها، تقع في نطاق جمهورية مصر العربية، كما تشترك جميعها في كونها تشترك في فهرس مركزي موحد موقعه الرئيس بالمجلس الأعلى للجامعات المصرية، والذي تتاح فرص وصلاحيات الدخول إليه – وفق أسس وقواعد وضوابط متفق عليها من وحدة المكتبة الرقمية – من قبل جميع المكتبات الأعضاء، للقيام بالمهام والأنشطة المختلفة المقررة والمتفق عليها من قبل إدارة مشروع الفهرس الموحد لاتحاد المكتبات الجامعية المصرية، وقد تم معالجة جوانب هذه الدراسة على مستوى هذه المكتبات على اختلاف أماكنها ومواقعها.

لجأ الباحث لإتمام الجوانب المختلفة لدراسته إلى التواصل مع تلك المكتبات والمؤسسات في أماكنها المختلفة، عبر وسائل التواصل المتعددة سواء (بالذهاب إلى بعضها والمقابلة بشكل مباشر مع المسؤولين – أو عبر البريد الإلكتروني – أو الاتصال التليفوني – أو الدخول على صفحات ومواقع تلك المؤسسات على الإنترنت).

فيما يتعلق بدراسة وتقييم نقاط الإتاحة الموضوعية المقننة بالتسجيلات الببليوجرافية عينة الدراسة فقد قام الباحث بالدخول على موقع الفهرس الموحد لاتحاد المكتبات الجامعية المصرية والحصول على العينة العشوائية المطلوبة وفق الأسس والمعايير المتفق عليها لسحب العينات.

كما تجدر الإشارة إلى أن التسجيلات الببليوجرافية الموجودة على موقع الفهرس الموحد لاتحاد المكتبات الجامعية المصرية تعكس أنشطة جميع المكتبات المنضمة إلى عضوية الاتحاد، حيث تشارك وتساهم جميع المكتبات الأعضاء في إدخال وإنشاء التسجيلات الببليوجرافية، كما تمتلك المكتبة صلاحيات التعديل على تلك التسجيلات الببليوجرافية وفق أسبقية الإدخال والاعتماد.

بناءً على ما سبق فإنه تمنح صلاحيات التعديل على التسجيلة الببليوجرافية فقط للمكتبة التي قامت بإدخالها أولًا، وتمنع تلك الصلاحيات من المكتبات التي قامت بإضافة نسخ فقط للوعاء الذي تمثله تلك التسجيلة، ويسمح فقط لتلك المكتبات تقييم تلك التسجيلة، ومن ثم وإرسال الملاحظات والتعديلات إذا لزم الأمر إلى المكتبة التي قامت بإدخال تلك التسجيلة.

❖ <u>الحدود الزمنية</u> :

منذ بداية العمل بمشروع المكتبة الرقمية وإنشاء اتحاد مكتبات الجامعات المصرية تحت مظلة المجلس الأعلى للجامعات في أوائل عام (2007 وحتى 2014).

❖ <u>الحدود اللغوية</u> :

تقوم الدراسة على تحليل عينة من التسجيلات الببليوجرفية العربية لأوعية المعلومات بالمكتبات الأعضاء بالفهرس الموحد لاتحاد المكتبات الجامعية المصرية والمتاحة من خلال موقع الفهرس الموحد لاتحاد المكتبات الجامعية المصرية على الإنترنت، حيث أن التسجيلات الببليوجرافية للأوعية باللغات الأجنبية، يتم استيرادها من فهارس المكتبات العالمية، وترد بها نقاط الإتاحة الموضوعية المقننة جاهزة.

❖ <u>الحدود النوعية</u>:

تقتصر الدراسة على نوع واحد من أنواع أوعية المعلومات، المتاحة على الفهرس الموحد، ألا وهو الكتب، حيث أن التسجيلات الببليوجرافية للكتب تمثل حوالي (80 %) من التسجيلات الببليوجرافية المتاحة على الفهرس الموحد لاتحاد المكتبات الجامعية المصرية.

<u>7/0 عينة الدراسة</u> :

قام الباحث بسحب عينة عشوائية من التسجيلات الببليوجرافية للكتب العربية الموجودة على الفهرس الموحد لاتحاد المكتبات الجامعية المصرية على الخط المباشر.

<u>8/0 حجم العينة</u> :

تم حساب حجم العينة بناءً على معادلة حجم العينة؛ لتقدير خاصية بالمجتمع للمجتمعات الكبيرة وتتمثل المعادلة فيما يلي :

$$n = \frac{Z_{\frac{\alpha}{2}}^2 \hat{p}(1-\hat{p})}{\varepsilon^2}$$

حيث إنَّ

- $n$ حجم العينة المطلوبة
- $\alpha = 0.05$ وذلك للحصول على ثقة (95%)
- $Z_{\frac{\alpha}{2}}^2$ هي القيمة الحرجة من جداول التوزيع الطبيعي المعياري، وتساوي (1.96) عند $\alpha = 0.05$.
- $\hat{p}$ تقدير أولي للخاصية بالمجتمع ويمكن وضعها عند (0.5)؛ لتعطي أكبر حجم عينة ممكن.
- $\varepsilon$ هامش خطأ في التقدير، وتم وضعه عند (0.03)؛ للحصول على دقة أعلى في النتائج.

استنادًا إلى ما سبق فإن حجم العينة الكلي المطلوب هو:

$$\frac{1.96^2 * 0.5 * (1 - 0.5)}{0.03^2} = 1068$$

باستخدام معادلة تصحيح حجم العينة فإنَّ:

$$n = \frac{1068}{1} + \frac{1 - 1068}{561650} = \frac{1068}{1.0019} = 1065.975$$

وبالتقريب يكون حجم العينة حوالي 1068 تقريبًا.

ومن ثم تم تقسيم الحجم الكلي للعينة على المجتمعات الجزئية، والممثلة بأفرع المعارف الانسانية وعددها (10) أفرع وذلك باستخدام استراتيجية التخصيص المتساوي، وعليه فإن حجم العينة المطلوبة من كل فرع من أفرع المعارف الانسانية هو:

$$n = \frac{1068}{10} = 107.$$

تجدر الإشارة إلى أنَّ الدراسة اعتمدت على سحب عينة عشوائية (Random Sample) من الشاملة "المجتمع" والتي بلغ عددها (561650) تسجيلة ببليوجرافية عربية والتي تعرف إحصائيًا بالشاملة الكبيرة "population" حيث بلغ عدد مفردات العينة (1068) تسجيلة ببليوجرافية بما نسبته (0.2%) تقريبًا من الحجم الكلي للشاملة، وهي نسبة مقبولة إحصائيًا، وتتناسب مع كل من التوزيع الطبيعي المعياري وحجم الشاملة، ويتناول الباحث في الفصل الرابع من الدراسة حجم العينة ومجتمع الدراسة بشكل أكثر تفصيلًا.

0/9 منهج الدراسة وأدوات جمع البيانات:

"المنهج هو الطريقة التي يتبعها الباحث في دراسته للمشكلة لاكتشاف الحقيقة، أو لتحقيق الهدف الذي قصد إليه من إعداد البحث"[1] ولأغراض الدراسة ومتطلباتها وأهدافها فقد استعان الباحث

---

[1] – عبد الهادي، محمد فتحي. (2013). **البحث ومناهجه في علم المكتبات والمعلومات**. (ط4). القاهرة: الدار

بأكثر من منهج لتحقيق تلك الأهداف، وتتمثل فيما يلي:

■ <u>المنهج الوصفي التحليلي:</u>

استعان الباحث بالمنهج الوصفي التحليلي؛ لدراسة رؤوس الموضوعات بالعينة العشوائية من التسجيلات الببليوجرافية للكتب العربية بالفهرس الموحد لاتحاد المكتبات الجامعية المصرية، ومن ثم فحصها وتحليلها طبقًا لمجموعة من الأسس والقواعد التي تحكم عملية التحليل الموضوعي، وتقنين رؤوس الموضوعات، ووفق الأدوات المعتمدة للتحليل الموضوعي بفهرس اتحاد المكتبات الجامعية المصرية.

■ <u>المنهج التقييمي:</u>

"عادة ما يطبق البحث التقييمي في الحالات التالية:

1- الأنشطة أو العمليات التي تجري في المكتبات ومراكز المعلومات.

2- نظام استرجاع معلومات المكتبة.

3- الأدوات الفنية للعمل مثل قوائم رؤوس الموضوعات ونظم التصنيف والمكانز.

4- مجموعات المكتبات من الكتب والدوريات والمواد السمعية والبصرية.. إلخ.

5- مصادر المعلومات المرجعية، مثل: دوائر المعارف والمعاجم اللغوية وكتب التراجم وأدلة الهيئات، إلخ".[1]

عليه فقد استعان الباحث بالمنهج التقييمي؛ لدراسة وتقييم الإمكانات الفعلية للقوائم والمكانز والأدلة الإرشادية التي تم الاستناد إليها في وضع نقاط الإتاحة الموضوعية المقننة بالتسجيلات الببليوجرافية؛ لتحديد درجة الجودة في هذه القوائم والمكانز والأدلة، ومدي كفايتها للتحليل الموضوعي لأوعية المعلومات بمكتبات الاتحاد. بالإضافة إلى قياس مدي الدقة في اختيار وصياغة رؤوس الموضوعات لتوفير نقاط الإتاحة الموضوعية المقننة وفق المعايير والسياسات التي أوصت بها وحدة المكتبة الرقمية بالمجلس الأعلى للجامعات وذلك من خلال قائمة المراجعة، والتي اشتملت على نقاط الإتاحة الموضوعية المقننة بالتسجيلات الببليوجرافية.

قام الباحث بإعداد مجموعة من الأدوات المهمة؛ لجمع البيانات يشار إليها فيما يلي:

<u>الاستبانة (1):</u>

تهدف إلى التعرف على القائمين بضبط جودة نقاط الإتاحة الموضوعية بالتسجيلات الببليوجرافية بالفهرس والمسؤولين عن وحدة المكتبة الرقمية بالمجلس الأعلى للجامعات المصرية (الجهة المشرفة على تنفيذ مشروع الفهرس)، والتعرف على مهام أعضاء فريق العمل بالوحدة، وما هي المهام والإجراءات والوظائف التي يقومون بها من أجل الارتقاء بجودة التسجيلات الببليوجرافية

---

المصرية للبنانية، ص 93.

[1] - نفس المرجع السابق، ص ص 141 : 142.

بصفة عامة ونقاط الإتاحة الموضوعية بصفة خاصة.

بالإضافة إلى التعرف على الصعوبات والمعوقات، التي قد تواجههم، وخططهم الحالية والمستقبلية لتحسين كفاءة استخدام أدوات التحليل الموضوعي، وقوائم رؤوس الموضوعات والمكانز؛ لمساعدة المكتبات الأعضاء في إنشاء تسجيلات ببليوجرافية ذات رؤوس موضوعات مقننة بدرجة عالية من الدقة في ظل التطورات المتلاحقة والمتسارعة في ظل البيئة الإلكترونية، وخاصة مع ظهور المعيار الجديد لوصف وإتاحة المصادر (وام RDA)؛ ولتحقيق هذا الهدف تم توجيه الاستبانة إلى القائمين على إدارة وحدة المكتبة الرقمية بالمجلس الأعلى للجامعات.

تتكون هذه الاستبانة من **تسعة محاور تشتمل على ثلاثين عنصرًا** تتعلق بأنشطة ومهام وبيئة العمل، بوحدة المكتبة الرقمية بالمجلس الأعلى للجامعات وهي:

- **فريق العمل بوحدة المكتبة الرقمية بالمجلس الأعلى للجامعات**، وقد اشتمل على عنصر واحد.
- **ضبط الجودة بوحدة المكتبة الرقمية**، وقد اشتمل على أربعة عناصر.
- **متابعة تحديثات مارك 21** وقد اشتمل على عنصرين.
- **التدريب والتأهيل**، وقد اشتمل على أربعة عناصر.
- **أدوات العمل الفنية المعتمدة بمشروع الفهرس الموحد لاتحاد المكتبات الجامعية المصرية للضبط الاستنادي للموضوعات**، وقد اشتمل على أربعة عناصر.
- **التعاون مع الجهات المحلية والدولية**، وقد اشتمل على أربعة عناصر.
- **العقبات والصعوبات**، وقد اشتمل على خمسة عناصر.
- **الخطط والتوجهات المستقبلية**، وقد اشتمل على عنصرين.
- **التوزيع العددي والنوعي للتسجيلات الببليوجرافية بالفهرس الموحد لاتحاد المكتبات الجامعية المصرية**، وقد اشتمل على خمسة عناصر.

قام الباحث بعدة خطوات لتحكيم الاستبانة قبل تطبيقها وهي:

- التواصل مع المسؤولين عن وحدة المكتبة الرقمية بالمجلس الأعلى للجامعات بالمقابلة الشخصية تارةً وبالتواصل عبر البريد الإلكتروني والهاتف تارةً أخرى، لتحكيم أدوات جمع البيانات ومن بينها هذه الاستبانة.
- عرض الاستبانة على مجموعة من أعضاء هيئة التدريس المهتمين بالتحليل الموضوعي والنظم الآلية وصيغ الاتصال سواء بالمقابلة الشخصية أو عبر إرسالها بالبريد الإلكتروني.
- عرض الاستبانة أيضًا على بعض الزملاء العاملين بفريق ضبط الجودة بوحدة المكتبة الرقمية بالمجلس الأعلى للجامعات.
- أسفرت الخطوات السابقة فيما يتعلق بتحكيم الاستبانة عن حذف بعض الأسئلة وتعديل أو

تغيير صياغة البعض الآخر، فضلًا عن إضافة أسئلة جديدة.

الاستبانة رقم (2):

تهدف إلى التعرف على القائمين بالتحليل الموضوعي لأوعية المعلومات (المتخصصين وغير المتخصصين) بالفهرس الموحد لاتحاد المكتبات الجامعية المصرية، والتعرف على الممارسات والإجراءات التي يقومون بها للتحليل الموضوعي لأوعية المعلومات، والتعرف على المهارات والخبرات التي تم اكتسابها من أجل القيام بذلك، ومعرفة أدوات التحليل الموضوعي، ومدى كفاءتها وكفايتها والتي يستعينون بها لإتمام مهامهم، بالإضافة إلى التعرف على الصعوبات والمعوقات التي قد تواجههم عند القيام بهذه العملية، ومقترحاتهم لتحسين كفاءة استخدام أدوات التحليل الموضوعي، وقوائم رؤوس الموضوعات؛ لإنشاء تسجيلات ببليوجرافية ذات رؤوس موضوعات مقننة بدرجة عالية من الدقة.

تجدر الإشارة إلى أن الاستبانة اشتملت على سبعة محاور تتضمن ثلاثة وأربعون عنصرًا يركز كل منها على إبراز الجوانب المختلفة للإجراءات والممارسات والأدوات المتعلقة بعملية التحليل الموضوعي لأوعية المعلومات، التي تتم من قبل العاملين بالمكتبات الأعضاء بفهرس اتحاد المكتبات الجامعية المصرية، وتمثلت محاور الاستبانة فيما يلي:

أ. **خصائص مجتمع الدراسة**. وقد اشتمل هذا المحور على سبعة عناصر.

ب. **قوائم رؤوس الموضوعات والمكانز**. وقد اشتمل هذا المحور على أحد عشر عنصرًا.

ت. **الإعداد والتأهيل**. وقد اشتمل هذا المحور على ستة عناصر.

ث. **التعامل مع المرافق الببليوغرافية**. وقد اشتمل هذا المحور على أربعة عناصر.

ج. **الممارسة والتطبيق**. وقد اشتمل هذا المحور على عنصرين.

ح. **المساعدة والإرشاد**. وقد اشتمل هذا المحور على ستة عناصر.

خ. **المشكلات والصعوبات**. وقد اشتمل هذا المحور على سبعة عناصر.

كما تم تحكيم الاستبانة أيضًا من قبل مجموعة[1] من أعضاء هيئة التدريس بالإضافة إلى بعض الخبراء العاملين في مجال المكتبات والمعلومات وخاصة فيما يتصل بالتحليل الموضوعي والضبط الاستنادي كما تم عرض الاستبانة أيضًا على بعض أعضاء هيئة التدريس المتخصصين في مجال الإحصاء والدراسات الإحصائية.

هذا وقد أفادت توجيهات السادة أعضاء هيئة التدريس في مجال المكتبات والمعلومات، بالإضافة إلى السادة أعضاء هيئة التدريس في مجال الإحصاء، وأيضًا الخبراء المتمرسين في مجال التحليل الموضوعي والضبط الاستنادي في إخراج الشكل النهائي للاستبانة بعد إجراء

التعديلات اللازمة من حذف لبعض الأسئلة، أو تعديل صياغة بعض الفقرات، أو تحديد عدد مناسب من الخيارات في حالة الأسئلة المغلقة، أو إضافة أسئلة جديدة تبدو ضرورية؛ لتغطية الجوانب المختلفة للموضوع.

تجدر الإشارة إلى أن هذه الاستبانة قد تم إعدادها باستخدام أحد خدمات (Google) التي يطلق عليها (Google form) أو (قالب جوجل)؛ مما سمح للباحث بتجربتها من خلال إرسال رابط الاستبانة عن طريق البريد الإلكتروني ومن خلال الرسائل الخاصة على موقع التواصل الاجتماعي فيس بوك (Facebook) إلى مجموعة من العاملين بمشروع فهرس اتحاد المكتبات الجامعية المصرية والذين يعملون في جامعات مختلفة، وقد أسفر هذا الإجراء عن استقبال عدد من الملاحظات والتعديلات من قبل المجموعة المشاركة والتي تركزت في جانبين هما:

الجانب الأول: فني أو تقني، يتعلق بتعديل بعض الإعدادات الخاصة بخدمة "قالب جوجل" (Google form) خاصة فيما يتعلق بضرورة تثبيت بعض الأسئلة في شكل إجباري أي لن يتمكن العضو المشارك من إرسال الاستبانة حتى يتم الإجابة عن هذا السؤال، كما تم تحويل بعض الأسئلة من خاصية إجباري إلى خاصية اختياري خاصة فيما يتعلق بالبيانات الشخصية للعضو كالاسم الثلاثي والبريد الإلكتروني...، إلخ.

الجانب الثاني: تخصصي، ركز على ضرورة إعادة صياغة بعض الأسئلة، وحذف البعض الآخر دفعًا للملل وتسهيلًا على المشاركين في الإجابة على أسئلة الاستبانة بما لا يخل بالمحتوي العام لها، بالإضافة إلى دمج بعض الأسئلة واختصار وتوضيح البعض الآخر.

<u>قائمة مراجعة بنقاط الإتاحة الموضوعية المقننة بالتسجيلات الببليوجرافية[1]</u> :

استعان الباحث بهذه القائمة كأداة لتقييم نقاط الإتاحة الموضوعية المقننة، والتعرف على توزيعها النوعي والعددي، وتكراراتها بالتسجيلات الببليوجرافية وفق الأقسام العشرة للمعرفة البشرية، بالإضافة إلى التعرف على الأخطاء الواردة بها، من حيث الشكل التي صيغت فيه، وهو قالب مارك للفهرسة المقروءة آليًا، ومن ناحية المضمون من خلال مضاهاتها بمضامين أوعية المعلومات التي تم فهرستها.

كما استعان الباحث بهذه الأداة للتعرف على مدى تقنين نقاط الإتاحة الموضوعية المقننة بالتسجيلات الببليوجرافية، من خلال مضاهاتها بقائمة رؤوس الموضوعات المعتمدة، وبالأدلة الإرشادية المقررة لضبط جودة نقاط الإتاحة الموضوعية المقننة.

كان إعداد هذه القائمة من الخطوات المهمة التي مرت بها الدراسة، وخاصة أنها تتناول نقاط الإتاحة الموضوعية المقننة لتقييمها وقياس جودتها وقد استعان الباحث في إعداد تلك القائمة

---

[1]- انظر الملحق رقم (4) قائمة مراجعة بنقاط الإتاحة الموضوعية المقننة بفهرس اتحاد المكتبات الجامعية المصرية.

بما يلي:

- الخبرة السابقة للباحث في العمل كاستشاري لضبط الجودة بمشروع الفهرس الموحد لاتحاد المكتبات الجامعية المصرية.
- القراءة المتأنية لأدب الموضوع المتعلق بصيغ الاتصال وخاصة صيغة مارك الببليوجرافي والاستنادي.
- الاطلاع المباشر على أحدث التعديلات وآخرها، التي تمت على الحقول المتعلقة بنقاط الإتاحة الموضوعية المقننة على موقع مكتبة الكونجرس الأمريكية https://www.loc.gov/ .
- المراجعة الكاملة والدقيقة للأدلة الإرشادية الصادرة عن وحدة المكتبة الرقمية بالمجلس الأعلى للجامعات، وخاصة فيما يتعلق بحقول الإتاحة الموضوعية المقننة.
- الدخول إلى النظام الآلي المسمى بنظام المستقبل لإدارة المكتبات (FLS)، وهو النظام المستخدم بمشروع الفهرس الموحد لاتحاد المكتبات الجامعية المصرية، والاطلاع على حقول الإتاحة الموضوعية الرئيسة والفرعية والمؤشرات بالتسجيلات الببليوجرافية والتعامل معها من خلال النوافذ المختلفة والإمكانات التي يتيحها النظام الآلي.
- الاطلاع على الدراسات السابقة في المجال والإفادة منها؛ لاستكمال بعض الجوانب والعناصر المتعلقة بالموضوع.

اعتمدت قائمة المراجعة على أحد خدمات (Google) التي يطلق عليها (Google form) أو (قالب جوجل) حيث يمكن من خلال تلك الخدمة الحصول على إحصاء وصفي للبيانات بالإضافة إلى إتاحة الحصول على تلك البيانات وتصديرها في أكثر من صيغة من الصيغ وخصوصًا في صيغة ملفات الجداول الإلكترونية (Excel sheet)، ومن ثم قام الباحث بترميز المتغيرات التي اشتملت عليها قائمة المراجعة وتحليل الأرقام الواردة باستخدام الحزمة الإحصائية (SPSS).

تجدر الإشارة إلى أن قائمة المراجعة تضمنت ثمانية محاور واشتملت تلك المحاور على (64) عنصرًا تمثلت في:

- **المحور الأول : (كود التسجيلة)** وقد اشتمل على عنصر واحد حيث تم وضع كود (رقم مسلسل) لكل تسجيلة بعد طباعتها من (1 : 107) ، موزعة وفق الأقسام العشرة للمعارف البشرية، حيث تم إدخال نتائج تقييم نقاط الإتاحة الموضوعية المقننة لكل تسجيلة ببليوجرافية على حدى في إطار القسم الذي تنتمي إليه تلك التسجيلة، أو بطريقةٍ أخرى فقد تم تقسيم عينة التسجيلات الببليوجرافية والتي بلغ عددها (1068) تسجيلة وفق الأقسام العشرة للمعارف البشرية إلى عشرة أقسام، بحيث يصبح لكل قسم (107) تسجيلة

ببليوجرافية ، ومن ثم تم ترميز كل تسجيلة في كل قسم من الأقسام العشرة من (1 : 107) ، وإدخال نتائج التقييم لكل التسجيلات في كل قسم على حدى، وفائدة هذا الترميز تكمن في تسهيل عملية الرجوع إلى التسجيلة الببليوجرافية، وتدارك الخطأ أو النسيان أو الخلط الذي قد يحدث أثناء إدخال البيانات، فضلًا عن تمييز نتائج تقييم كل قسم على حدى لأغراض المقارنة.

- **المحور الثاني: أقسام المعارف البشرية** وقد اشتمل على عنصر واحد حيث يتم اختيار القسم الذي تتبعه التسجيلة الببليوجرافية وقد سبق توضيح ذلك.

- **المحور الثالث: إجمالي رؤوس الموضوعات بالتسجيلة**، اشتمل على عنصر واحد وقد تم وضع خانات الاختيار (من 1: 5 فأكثر)

- **المحور الرابع: نقطة الإتاحة الموضوعية المقننة رأس موضوع اسم شخص (600)** وقد اشتمل هذا المحور على أحد عشر عنصرًا تمثل خصائص نقطة الإتاحة الموضوعية المقننة (رأس موضوع اسم شخص) وفيما يلي عرض لتلك الخصائص وقد تكررت في باقي نقاط الإتاحة الموضوعية المقننة بقائمة المراجعة:

  ◼ 1/4 توفر التاج وقد تم وضع الاختيار وفق الإجابة (نعم أو لا) والتي تم ترميزها وفق الحزمة الإحصائية وفق الرقمين (1، 2) على التوالي لأغراض التحليل الإحصائي.

  ◼ 2/4 تكرار التاج وقد وضع الاختيار بين الحد الأدنى (1) إلى الحد الأقصى (5 فأكثر).

  ◼ 3/4 المؤشرات.

  ◼ 4/4 التفريعات.

  ◼ 5/4 الحقول الفرعية.

  ◼ 6/4 ترتيب الحقول الفرعية لنقاط الإتاحة الموضوعية المقننة.

  ◼ 7/4 مصدر وموقع رأس الموضوع وفق القالب مارك.

  ◼ 8/4 المحتوى.

  ◼ 9/4 صياغة رؤوس الموضوعات

  ◼ 10/4 علامات الترقيم والمسافات الطباعية.

  ◼ 11/4 الأخطاء اللغوية والإملائية.

- **المحور الخامس: نقطة الإتاحة الموضوعية المقننة رأس موضوع اسم هيئة (610)** وقد اشتمل هذا المحور على أحد عشر عنصرًا تمثل خصائص نقطة الإتاحة الموضوعية المقننة رأس موضوع اسم هيئة.

- **المحور السادس: نقطة الإتاحة الموضوعية المقننة رأس موضوع عنوان تقليدي أو موحد (630)** وقد اشتمل هذا المحور على ثلاثة عشر عنصرًا تمثل خصائص نقطة الإتاحة الموضوعية المقننة رأس موضوع عنوان تقليدي أو موحد.
- **المحور السابع:** نقطة الإتاحة الموضوعية المقننة رأس موضوع مصطلح موضوعي (650) وقد اشتمل هذا المحور على ثلاثة عشر عنصرًا تمثل خصائص نقطة الإتاحة الموضوعية المقننة رأس موضوع مصطلح موضوعي.
- **المحور الثامن: نقطة الإتاحة الموضوعية المقننة رأس موضوع اسم جغرافي (651)** وقد اشتمل هذا المحور على ثلاثة عشر عنصرًا تمثل خصائص نقطة الإتاحة الموضوعية المقننة رأس موضوع اسم جغرافي.

قام الباحث بنفسه، وبالاستعانة ببعض الخبراء بتجريب قائمة المراجعة في الشكل الإلكتروني وذلك بتجربة إدخال قسم كامل من أقسام المعارف البشرية، والذي يتضمن تقييم لعدد (107) تسجيلة، والتعرف على أهم الأخطاء والمشاكل التي تحدث أثناء إدخال البيانات؛ لضمان أقصى درجات الدقة في إدخال البيانات، ومن ثَمَّ سلامة الملفات الناتجة من علمية الإدخال من الأخطاء، كما قام الباحث أيضًا بتصدير ملفات الجداول الإلكترونية ومضاهاتها بالبيانات الواردة بها مع النسخ المطبوعة للتسجيلات الببليوجرافية؛ للتأكد من إدخال جميع التسجيلات الببليوجرافية دون استثناء.

10/0 المصطلحات المرتبطة بالدراسة:

تردد في الدراسة عدد من المصطلحات والمفاهيم، بعضها قديم، والبعض الآخر مستحدث في المجال، مما يستدعي سردها وبيان مضامينها، إما لتحديد معناها في إطار الدراسة، أو إزالة الغموض والالتباس عنها في حالة ملامستها وتداخلها مع مصطلحات أخرى، "وليس هناك من شك بأن تحديد المصطلحات والمفاهيم تعتبر إحدى الطرق المنهجية الهامة في العلوم الإنسانية والاجتماعية، لأن لغتهما مجردة كما هي حال لغات العلوم الأخرى، وأن مدلول كل مصطلح يتحدد بموضوع البحث الذي هو مرتبط به"[1].

■ الفهرس الموحد: Union catalog

تعريف الشامي "فهرس واحد يشتمل على جميع المواد، التي قد تكون موجودة في موقع واحد، أو عدة مواقع وأماكن مختلفة، مع تحديد مواقعها، وقد يكون فهرسًا بالمؤلفين، أو بالموضوعات لجميع المواد، أو مختارات منها، وقد يكون محصورًا في موضوعات معينة، أو نوع

---

[1] - المغربي، كامل محمد. (2011). **أساليب البحث العلمي في العلوم الإنسانية والاجتماعية**. (ط.4). عمان: دار الثقافة للنشر والتوزيع، ص 45.

معين من المطبوعات. ومن الأمثلة على الفهرس المباشر الموحد (WorldCat)"[1].

وجاء في قاموس علم المعلومات والمكتبات على الخط المباشر أودليس (ODLIS) (Online Dictionary for Library and Information Science) أنه "قائمة بمقتنيات كافة المكتبات في نظام المكتبة، أو كل، أو جزء من مجموعات مجموعة من المكتبات المستقلة، مشيرًا بالاسم و / أو رمز موقع المكتبة التي تمتلك على الأقل نسخة واحدة من العنصر. عندما يكون الغرض الرئيس للفهرس الموحد هو الإشارة إلى الموقع، فإن الوصف الببليوجرافي المتوفر لكل مدخل، يمكن تخفيضه إلى أدنى حد ممكن، ولكن عندما يعمل أيضا لأغراض أخرى، فإن الوصف يكون أكثر اكتمالاً. وعادة ما يكون ترتيب الفهرس الموحد ترتيبًا أبجديًا بالمؤلف، أو العنوان"[2].

■ الفهرس الافتراضي الموحد: (Virtual Union Catalog)

"نظام آلي للبحث في مقتنيات اثنين، أو أكثر من فهارس المكتبات المنفصلة بعضها عن بعض، وذلك باستخدام بروتوكول (Z39.50) و / أو آليات أخرى للبحث واسترجاع المعلومات، على النقيض من الفهرس المركزي الموحد، والذي يتم فيه جمع تسجيلات الفهرس في قاعدة بيانات واحدة، أو موقع مادي واحد"[3].

■ قاعدة البيانات: (Database)

"هي مجموعة من البيانات المهيكلة المقيدة بالحاسب الآلي، وخاصة تلك التي يمكن الوصول إليها بطرق مختلفة:"[4]، أو "ملف كبير بالمعلومات الرقمية (التسجيلات الببليوجرافية، والملخصات، وثائق النص الكامل، مداخل الدليل، والصور، والإحصاءات، الخ)، والتي يتم تحديثها بانتظام، والمتعلقة بموضوع أو مجال معين، وتتألف من شكل موحد من التسجيلات نظمت؛ لسهولة وسرعة البحث والاسترجاع بمساعدة برمجيات نظام إدارة قواعد البيانات (DBMS)"[5].

■ نظام إدارة قاعدة البيانات: (DBMS)

"تطبيق كمبيوتر مصمم للتحكم في خزن واسترجاع وأمن وسلامة البيانات، وإعطاء تقرير بالبيانات في شكل سجلات موحدة، نظمت في ملف ضخم قابل للبحث يدعى قاعدة بيانات. نطاق برمجيات إدارة قواعد البيانات (DBMS) يمتد من أنظمة بسيطة مخصصة لأجهزة الكمبيوتر

---

[1] - الشامي، أحمد محمد. (يناير 2014). **مصطلحات المكتبات والمعلومات والأرشيف**، تاريخ الوصول 4 أبريل 2015 في /http://www.elshami.com
[2] - Reitz, J. (2014, January 1). ABC-CLIO. Retrieved April 4, 2015, from http://www.abc-clio.com/ODLIS/odlis_u.aspx
[3] - Ibid.
[4] - Definition of database in English: (2015, January 1). Retrieved April 4, 2015, from http://www.oxforddictionaries.com/definition/english/database
[5] - Ibid.

الشخصية إلى أنظمة معقدة للغاية مصممة لتعمل على الحاسبات"[1].

■ الفهرس المباشر أو الفهرس المتاح للجمهور على الخط المباشر: Online Public Access Catalog (OPAC)

"هو قاعدة بيانات تتألف من تسجيلات ببليوجرافية، تصف الكتب وغيرها من المواد التي تمتلكها مكتبة ما، أو تجّمع مكتبات تتاح عبر منافذ عامة، أو محطات عمل تتركز في العادة بالقرب من مكتب المراجع بالمكتبة، حيث يمكن للمستفيدين أن يجدوا سهولة في طلب المساعدة من اختصاصي مراجع مدرب، ويمكن البحث في معظم الفهارس بالمؤلف، والعنوان، والموضوع، والكلمات المفتاحية. وتتيح للمستفيدين إمكانات الطباعة والتحميل وتصدير التسجيلات لحساب بريد إلكتروني"[2].

■ قائمة موحدة: Union list

قائمة (وعادة مع التسجيلات الببليوجرافية أقل اكتمالًا من الفهرس الموحد) بمقتنيات اثنين أو أكثر من المكتبات. وعادة ما تقتصر على نوع معين من الوثائق (الصحف والمخطوطات، والمواد عن مكان، أو موضوع، أو شخص معين، وهلم جرا)، ومجموعة محددة جيدًا من المكتبات أو المحفوظات (مثل تلك الموجودة في مدينة معينة أو بلد، أو من نوع معين)[3].

■ مفهوم الجودة :

تعرف الـ (**ASQ**) الجودة بأنها "مصطلح ذاتي حيث لكل شخص أو قطاع تعريفه الخاص به. ففي الاستخدام التقني، يمكن أن تشتمل الجودة على اثنين من المعاني:

(1) خصائص المنتج أو الخدمة التي تؤثر على قدرته على تلبية الاحتياجات المعلنة أو الضمنية.

(2) منتج أو خدمة خالية من العيوب"[4] وتحتفظ كل جهة وفق معاييرها وأسسها في تفسير مفهوم الجودة وفق هذا التعريف.

■ جودة التسجيلات الببليوجرافية :

يذخر أدب المكتبات بتعريفات عديدة لجودة الفهرسة أو جودة التسجيلات الببليوجرافية من أهم هذه التعريفات "التعريف الذي قدمته بربارا تيليت – رئيس مكتب السياسات ودعم الفهرسة في مكتبة الكونغرس– في منتدى الفهرسة في (17 أكتوبر 1994)، حيث عرفت جودة الفهرسة بأنها

---

[1]- Ibid.

[2]- عبد الهادي، محمد فتحي. (2010). **الفهارس العربية المتاحة على الخط المباشر والمعايير الببليوجرافية القياسية**. – مجلة الملك فهد الوطنية، مج 16، (ع 2)، ص 228.

[3] - Feather, J. (2003). International encyclopedia of information and library science (2nd ed., pp. 643 - 644). London: Routledge.

[4]- Quality Glossary - Q. (n.d.). Retrieved March 8, 2015, from http://asq.org/glossary/q.html

المعلومات الببليوغرافية الدقيقة التي تلبي احتياجات المستخدمين وتوفر الوصول المناسب في الوقت المناسب"[1]

- ■ <u>نقاط الإتاحة: Access Points</u>

نقاط الإتاحة عامة يتردد صداها في الفهارس المتاحة للجمهور على الخط المباشر وفي الفهارس اليدوية، ولكنها توفر الوصول الإضافي الذي يعزز إلى حد كبير استرجاعها. وإمكانية البحث عن طريق "الكلمة" أو "النص الحر"، يحرر المستخدم من ضرورة أن يكون لديه المعلومات الكاملة عن المؤلف، أو العنوان.

ومع ذلك فإن العديد من الفهارس على شبكة الإنترنت لا تتضمن الضبط الاستنادي الكامل للمؤلفين، والموضوعات، وهذا النقص يمكن أن يؤدي إلى نتائج غير دقيقة في البحث، خلافا لأفضل الممارسات في الفهارس اليدوية"[2].

ويمكن تعريف نقاط الإتاحة بأنها "الوحدة (الاسم أو المصطلح أو العبارة أو الرمز، الخ.) التي يُبحث تحتها عن التسجيلة الببليوجرافية ويُتَعرف عليها بها. وكأمثلة: المؤلفون، والعناوين، والموضوعات بالفهارس، وفي الفهارس البطاقية كانت نقاط الإتاحة المستخدمة تتضمن المدخل الرئيس والمداخل الإضافية ورؤوس الموضوعات، ورقم التصنيف، أو رقم الاستدعاء، ولكن بعد استخدام الفهارس المقروءة آليًا، أصبح كل جزءٍ تقريبًا في تسجيلات الفهارس (مثل اسم الناشر، ونوع المادة ... الخ.)[3].

- ■ <u>نقاط الإتاحة الموضوعية: Subject access points</u>

يعرف (Hjørland, B, 2015)[4] نقاط الإتاحة الموضوعية بأنها أجزاء قابلة للبحث من تمثيلة الوثيقة (أو وثيقة النص الكامل)، كما يمكن تعريف نقاط الإتاحة الموضوعية بمفهومها الواسع بأنها المصطلحات التكشيفية التي تعكس المحتوى الموضوعي للوعاء والتي أعدت خصيصًا لاستخدامها في البحث الموضوعي.

- ■ <u>نقاط الإتاحة الموضوعية المقننة: Authorized Subject access points (ASAPₛ)</u>

تحمل نقاط الإتاحة الموضوعية مفهومًا ضيقًا يمكن وصفه بأنه تلك المصطلحات المقيدة أو المقننة التي يتم اختيارها من قبل الاختصاصيين الموضوعيين (أو إخصائي التحليل الموضوعي) من أدوات معدة سلفًا كالمكانز وقوائم رؤوس الموضوعات لوصف المحتوى الموضوعي لأوعية

---

[1] - Thomas, S. (1996). Quality in Bibliographic Control. Retrieved August 29, 2015, from https://ecommons.cornell.edu/handle/1813/2669

[2] - Quality Glossary - Q. (n. d.). ,Op.Cit,.

[3] - الشامي، أحمد محمد. (يناير 2014). **مصطلحات المكتبات والمعلومات والأرشيف**، تاريخ الوصول 4 أبريل 2015 في http://www.elshami.com/

[4] - Hjørland, B. (2015, August 28). Subject access points. Retrieved from https://www.academia.edu/Messages/854440

المعلومات، وتسمى نقاط الإتاحة الموضوعية المقننة (ASAP$_s$) Authorized Subject access points.

■ الفهرسة الآلية أو المحوسبة: (automation cataloging)

يعرف (Taylor 2006) الفهرسة: بأنها عملية إنشاء تسجيلات بديلة لحزم المعلومات، عن طريق وصف حزمة المعلومات واختيار نقاط الإتاحة: الاسم والعنوان، وإجراء التحليل الموضوعي مع تحديد رؤوس الموضوعات وأرقام التصنيف وصيانة النظام حتى تصبح التسجيلات متاحه من خلاله"[1].

أما الفهرسة المحوسبة / المؤتمتة (automation cataloging) " فهي عملية إدخال ومعالجة واسترجاع البيانات الببليوجرافية ضمن الإطار العام للقواعد المقننة المعتمدة والمعمول بها، مع استثمار قدرات البرمجيات والحواسيب؛ لضمان منافذ أكثر عددًا ومرونة للمستخدمين والمستفيدين النهائيين معًا"[2].

**0/11 الدراسات السابقة:**

لاحظ الباحث من خلال مطالعته ندرة شديدة في الإنتاج الفكري العربي الذي يتناول رؤوس الموضوعات والتحليل الموضوعي ونقاط الإتاحة الموضوعية في الفهارس الآلية على الخط المباشر، خاصة في الأطروحات الأكاديمية أو الجامعية، بينما على الجانب الآخر شهد الإنتاج الفكري الأجنبي وفرة في الأبحاث التي تناولت هذا الموضوع، سواء في الكتب أو مقالات المجلات العلمية، كما أن هناك ندرة في الأطروحات الجامعية التي تتناول نقاط الإتاحة الموضوعية المقننة في الفهارس الآلية على الخط المباشر.

عبر البحث في قاعدة بيانات الرسائل الجامعية (UMI ProQuest dissertations and theses) في مطلع العام (2010) تحت المصطلح quality control & subject headings فقد أسفرت نتائج البحث عن رسالة واحدة ذات صلة بموضوع البحث، وبالبحث أيضًا في نفس المستودع تحت مصطلح (subject access points and online catalog) فقد أسفرت نتائج البحث عن رسالتين جامعتين فقط، وبتجديد البحث مرة أخرى في قاعدة البيانات في العام (2012) ، فقد حصل الباحث على رسالة واحدة حول مفهوم جودة الفهرسة.

فيما يلي عرض لأهم الدراسات والأبحاث المتعلقة بشكل مباشر بموضوع الدراسة.

---

[1]- Taylor, A.G. (2006). Introduction to cataloging and classification. Westport, CT: Libraries Unlimited.

[2] - قنديلجي، عامر إبراهيم، والسامرائي، إيمان فاضل. (2004). **حوسبة (أتمتة) المكتبات: استثمار إمكانات الحواسيب في إجراءات وخدمات المكتبات ومراكز المعلومات**. (ط1). عمان: دار المسيرة للنشر والتوزيع، ص 26.

## أولًا: الدراسات العربية:

### محمد فتحي عبد الهادي (1975) [1]

تعد هذه الدراسة من أوائل الدراسات في مجال التحليل الموضوعي ورؤوس الموضوعات وهي "تتضمن دراسة ميدانية للممارسات والتطبيقات في فهارس المكتبات والببليوجرافيات والكشافات، كما تتضمن وضع قواعد لرؤوس الموضوعات العربية، وتنتهي الرسالة بوضع أسس إعداد قائمة عربية في العلوم الاجتماعية فضلًا عن القائمة نفسها"[2].

### رفل نزار محمد علي قاسم (2005) [3]

ركزت الدراسة على تحليل رؤوس الموضوعات العربية والأجنبية، المستخدمة في الفهرس الموضوعي للمكتبة المركزية بجامعة الموصل في العلوم الهندسية؛ في محاولة لقياس أداء هذا الفهرس وقدرته على الإتاحة الموضوعية، وقد استخدمت في القياس بعض المعايير العالمية لاسيما معدل عدد رؤوس الموضوعات للكتاب الواحد وعدد الرؤوس المستخدمة في البطاقة الواحدة.

أعطت الدراسة أهمية قصوى لدراسة مدى التزام المكتبة المركزية بالمبادئ الأساسية لاختيار رؤوس الموضوعات وصياغتها، وتأثير الإخلال بهذه المبادئ على أداء الفهرس كما أُخذ تأثير الأخطاء الإملائية والأخطاء في علامات الترقيم بعين الاعتبار أيضًا عند القياس والتقويم.

استخدمت الدراسة المنهج الوصفي التحليلي في تحليل عينة رؤوس الموضوعات في العلوم الهندسية التي بلغت (395) رأس موضوع باللغة العربية و (588) رأس موضوع باللغة الإنكليزية تم إعدادها من عينة أُخرى من بطاقات الفهرس بلغت (614) بطاقة عربية و (1035) بطاقة أجنبية.

خلصت الدراسة إلى ضعف أداء الفهرس الموضوعي للمكتبة المركزية، وقد خرجت الدراسة ببعض المقترحات التي تهدف إلى تحسين أداء الفهرس الموضوعي للمكتبة المركزية لاسيما عن طريق الاستعانة بخبرة وأداء ومقترحات المتخصصين في مجال المعلومات والمكتبات، وتحديث قوائم رؤوس الموضوعات المستخدمة في المكتبة وتنظيم برامج تدريبية للعاملين في قسم الفهرسة.

### ليلى سيد سميع (2007) [4]

---

[1] - عبد الهادي، محمد فتحي. (1975). **إنشاء قائمة رؤوس موضوعات عربية في العلوم الاجتماعية**. – القاهرة، – 2 مج. – رسالة دكتوراه– جامعة القاهرة. كلية الآداب. قسم المكتبات والوثائق.

[2] - حافظ، سرفيناز محمد. (2009). **قوائم رؤوس الموضوعات العربية: دراسة تقييمية وتحليلية للقائمة الكبرى لرؤوس الموضوعات العربية**، مجلة الفهرست.

[3] - قاسم، رفل نزار. (2005). **رؤوس الموضوعات المستخدمة في المكتبة المركزية بجامعة الموصل في مجال العلوم الهندسية دراسة تحليلية**، أطروحة ماجستير، جامعة الموصل، كلية الآداب – قسم المكتبات والمعلومات.

[4] - سميع، ليلى سيد. (2007). **نظم الضبط الاستنادي الآلي للموضوعات: دراسة ميدانية للتطبيقات بالمكتبات**

استهدفت الدراسة التعرف على ممارسات واجراءات الضبط الاستنادي للموضوعات باستخدام النظم الآلية في المكتبات المصرية، والتعرف على الإمكانات والتطبيقات التي تقدمها النظم الآلية في عمليات الضبط الاستنادي الآلي لرؤوس الموضوعات واختارت الدراسة أربع مكتبات مصرية تقتني نظمًا آلية وهي: دار الكتب الوطنية، مكتبة الإسكندرية، المكتبة القومية الزراعية، مكتبة مبارك – مصر العامة الآن – وتوصلت الدراسة لعدة نتائج تتلخص فيما يلي:

- أن الاتجاه يسير الآن في مجال الضبط الاستنادي نحو بناء ملفات استناد متعددة اللغات؛ لزيادة فرص الإفادة لصالح المستفيد.
- تعاني القوى البشرية المخصصة لإعداد تسجيلات استناد مقروءة آليًا لرؤوس الموضوعات في المكتبات محل الدراسة من قصور في عمليات التأهيل والتدريب المقدم بخلاف العاملين بمكتبة الإسكندرية الذين تلقوا التدريب الكافي فضلًا عن الجهود الذاتية التي بذلت من جانب العاملين.
- لا توجد سياسات أو اجراءات مكتوبة، ومعلنة لمعدي التسجيلات الاستنادية حول ارشادات بناء الملف الاستنادي.
- هناك مشكلات تواجه العاملين عند إعداد التسجيلات الاستنادية باللغة العربية.

وخلصت الدراسة إلى مجموعة من التوصيات من بينها:

- ضرورة اهتمام أقسام المكتبات بالجامعات المصرية بمقرر الضبط الاستنادي الآلي للموضوعات.
- ضرورة اهتمام موردي النظام بعملية تهيئة النظم الفرعية للاستناد.
- ينبغي عند بناء الملف الاستنادي الوطني لرؤوس الموضوعات أن يتم تحديد جهة إشرافية تتولى عملية التنسيق والمتابعة والحصول على الدعم المادي من الجهات المهتمة بالأمر، والتسويق للمشروع كمتطلب وطني.

**خالد محمد عبد الفتاح (2011)** [1]

هدفت الدراسة إلى تقييم تجربة بناء الفهرس الموحد لمقتنيات المكتبات الجامعية المصرية كأحد التجارب الرائدة في المنطقة العربية لبناء فهارس موحدة للمكتبات الأكاديمية على المستوى الوطني، وركزت الدراسة بصفة أساسية على تأثير نموذج العمل بالفهرس الموحد على مكونات البنية التحتية للفهرس، والنظام الآلي المستخدم، وأساليب دمج البيانات في كل مرحلة من مراحل

---

**المصرية**، أطروحة دكتوراه، جامعة القاهرة، كلية الآداب – قسم المكتبات والوثائق والمعلومات.

[1] - عبد الفتاح، خالد محمد. (2011). **الفهرس الموحد للمكتبات الجامعية المصرية: دراسة حالة**، مجلة الفهرست، مج 9، ع 63، ص ص 29 – 105.

البناء موزعة على المكتبات المشاركة بقطاعاتها المختلفة، وإجراءات ضبط الجودة والتحكم في عمليات الإدخال ومعالجة المكررات، بالإضافة إلى القوى البشرية وآليات تشغيلها والتحديات وطرق مواجهتها.

خلصت الدراسة إلى أن النموذج المادي المركزي هو النموذج الملائم لحالة المكتبات المصرية التي تتوافق كثيرًا مع حالة العديد من المكتبات العربية، ومكتبات الدول النامية.

**طه نبيل عبد الحميد الفرماوي (2012)** [1]

هدفت الدراسة إلى دراسة وتحليل جودة التسجيلات العربية لأوعية معلومات مكتبات جامعية المنوفية من خلال أخذ عينة عشوائية من التسجيلات العربية بمكتبات الجامعة والبالغ عددها (13458)، والتي تم اعتمادها في الفترة ما بين (2007 – 2010) أي خلال ثلاث سنوات من بدء مشروع الفهرسة الآلية بمكتبات الجامعة، وقد خلصت الدراسة إلى ضعف التسجيلات الببليوجرافية بفهرس جامعة المنوفية.

**ثانيًا: الدراسات الأجنبية:**

**Mansor, Yushiana (1999)** [2]

تدور هذه الدراسة حول بحث إمكانيات استحداث برنامج للفهرسة التعاونية المشتركة على غرار برامج الفهرسة التعاونية الأخرى، والتي نجح تطبيقها في أمريكا الشمالية وأوروبا، وفى مناطق أخرى من العالم، ويرى الباحث أن واحدًا من أهم العوامل التي تحدد نجاح هذا البرنامج، هو التوافق الذي يتم به إعداد سجلات مارك للفهرسة التعاونية بين المكتبات المشاركة.

الغرض الأساس من هذا البحث، هو دراسة الاختلافات في ممارسات الفهرسة، ومدى تأثيرها على التوافق بين تسجيلات مارك في المكتبات الماليزية.

تم أخذ عينه مكونة من (410) تسجيلة ببليوجرافية من سجلات مارك تم استخراجها من قواعد بيانات الفهرس الموحد على الخط المباشر لثلاثة مكتبات جامعية بماليزيا، والجامعات هي: جامعة ماليزيا ساراواك (UNIMAS) University Malaysia Sarawak (الجامعة الأم)، وجامعة بيرتيان ماليزيا (UPM) University Pertanian Malaysia، وجامعة المالايا University Malaya (UM).

---

[1] - الفرماوي، طه نبيل عبد الحميد. (2012). **جودة التسجيلات العربية في فهرس مكتبات جامعة المنوفية: دراسة ميدانية**، أطروحة ماجستير، جامعة المنوفية، كلية الآداب، قسم المكتبات والمعلومات.

[2] - Mansor, Yushiana (1999). Issues in developing a cooperative cataloging program in Malaysia: An analysis of MARC records in three university libraries' OPAC databases. Ph.D. dissertation, University of Pittsburgh, United States -- Pennsylvania. Retrieved January 13, 2010, from Dissertations & Theses: Full Text. (Publication No. AAT 9928026).

تم تحليل كل تسجيلة من التسجيلات الببليوجرافية؛ لدراسة الاختلافات الموجودة (رؤوس الأسماء) و (رؤوس العناوين) و(رؤوس الموضوعات)، كما صنفت الاختلافات الى "اختلافات الشكل" و "اختلافات المحتوى" و "التحرير" و "اختلافات الادخال".

أشارت نتائج هذه الدراسة إلى وجود اختلافات في ممارسات الفهرسة بماليزيا، وتكمن هذه الاختلافات الرئيسة في شكل ومحتوى تسجيلات مارك، وتشير هذه النتائج إلى افتقار المفهرسين الماليزيين إلى الخبرة والكفاءة، وعدم إلمامهم بمعايير شكل الاتصال مارك، وقائمة مختصرات قواعد الفهرسة الأنجلو أمريكية، والوصف الببليوجرافي المعياري، وقائمة رؤوس موضوعات مكتبة الكونجرس الأمريكية، ويجب اتخاذ التدابير المناسبة، سواء على المستويات المحلية أو الوطنية؛ من أجل تبسيط، وتوحيد ممارسات الفهرسة المختلفة، والنجاح في دمج التسجيلات الببليوجرافية في مختلف قواعد البيانات، وجعلها في ملف واحد، بالإضافة إلى ذلك، بحث القيام بممارسة ضبط الجودة للحفاظ على قاعدة البيانات المشتركة ذات جودة عالية.

**Xu, Hong (1996)** [1]

الغرض من هذه الدراسة هو تحديد إلى أي مدى تتداخل حقول رقم التصنيف، والعنوان، ورؤوس الموضوعات مع بعضها البعض في التسجيلات الببليوجرافية بالفهرس على الخط المباشر؛ لتوفير نقاط الإتاحة الموضوعية؛ ولتحديد نقاط الإتاحة الموضوعية الفريدة الموجودة في هذه الحقول، ولاستكشاف هذه العلاقة بين الثلاثة حقول الغنية بالموضوع، وأربع فئات من العلوم البشرية بتصنيف ديوي العشري وهي:

300 العلوم الاجتماعية
500 العلوم البحتة (العلوم الطبيعية والرياضيات)
600 العلوم التطبيقية (التكنولوجيا)
700 الفنون (الفنون الجميلة والديكور)

تم اختيار عينه عشوائية من التسجيلات الببليوجرافية بلغت (205) تسجيلة ببليوجرافية، تمثل المنفردات المنشورة باللغة الإنجليزية، وذلك في الفترة ما بين (1990- 1994)، وقد تم انتقاء هذه العينة العشوائية من الفهرس الموحد على الخط المباشر لمركز دعم خدمات المكتبات بالحاسب الآلي (OCLC)، والمقاييس التي صممت لقياس التكرار، كانت تستخدم لاختبار ما إذا كانت هناك فروق ذات دلالة إحصائية في متوسط عدد نقاط الإتاحة الموضوعية التي تتوفر من تداخل

---

[1] - Xu, Hong (1996). Redundancy and uniqueness of subject access points in online catalogs. Ph.D. dissertation, University of Illinois at Urbana-Champaign, United States -- Illinois. Retrieved January 14, 2010, from Dissertations & Theses: Full Text.(Publication No. AAT 9712489).

حقول (رقم تصنيف، و العنوان) ، ( رقم التصنيف، و رؤوس الموضوعات)؛ ( العنوان، و رؤوس الموضوعات)، ومتوسط عدد نقاط الإتاحة الموضوعية الفريدة التي توفرها حقول ( رقم التصنيف والعنوان ، و رؤوس الموضوعات) ، أو حقول ( رقم التصنيف / العنوان)، ( رقم التصنيف / رؤوس الموضوعات ، والعنوان / رؤوس الموضوعات )، وأي قطاع من قطاعات المعارف البشرية الأربعة السابقة، أكثر تداخلًا أو انفرادًا في نقاط الإتاحة الموضوعية، وأي توزيع أكثر تداخلًا وانفرادًا لنقاط الإتاحة الموضوعية، هل في ثلاثة أو كل زوج من الحقول يعتمد على أربعة أقسام من العلوم الواردة بتصنيف ديوي العشري.

وجد أنه في التسجيلات الببليوجرافية بالفهرس، متوسط عدد نقاط الإتاحة الموضوعية المتوفرة من تداخل حقول (رقم التصنيف، والعنوان) هي أقل بكثير من نقاط الإتاحة الموضوعية المتوفرة من تداخل حقول (رؤوس الموضوعات/ رقم التصنيف)، وحقل (العنوان/ رؤوس الموضوعات)، ومتوسط عدد نقاط الإتاحة الموضوعية الناتجة من التداخل الناتج من الثلاث تقسيمات أو في كل زوج من التقسيمات في كل قطاع من قطاعات المعرفة البشرية.

تبين أيضًا أنَّ متوسط عدد نقاط الإتاحة الموضوعية هو الأعلى في حقل رؤوس الموضوعات وأدنى في حقل العنوان ؛ كما أن الأربعة أقسام من المعارف البشرية تختلف اختلافًا كبيرًا في أعداد نقاط الإتاحة الموضوعية الفريدة وأغلبها في قطاع (300) العلوم الاجتماعية وأقلها في قطاع (500) العلوم البحتة أو الطبيعية وهذا التوزيع يعنى أن متوسط أعداد نقاط الإتاحة الموضوعية الفريدة المتوفرة من تداخل حقول (رقم التصنيف والعنوان) و حقل رؤوس الموضوعات أو ( رقم التصنيف والعنوان) أو ( العنوان ورؤوس الموضوعات) مرتبطة بشدة بحقول رؤوس الموضوعات والعنوان ومرتبطة بشدة بالقطاعات المعرفية وأن حقل رؤوس الموضوعات يعد المصدر الرئيس لنقاط الإتاحة الموضوعية ونقاط الإتاحة الموضوعية الفريدة، وبالرغم من ذلك على المستوى النظري فإن إضافة أفكار موضوعية من حقل رقم التصنيف و/أو العنوان هو الأفضل لتسهيل الاسترجاع أو بالأحرى أدق في الاستدعاء.

**Schoenung, J. (1981)** [1]

كان الغرض من هذا البحث تقييم فعالية النظام الفرعي للفهرسة المشتركة بشبكة المكتبات المحوسبة على الخط المباشر (OCLC) مع التركيز بوجه خاص على جودة التسجيلات الببليوجرافية للكتب التي يتم إدخالها عن طريق المكتبات الأعضاء.

كانت المنهجية بأخذ عينة عشوائية مكونة من (1107) تسجيلة ببليوجرافية من التسجيلات التي يتم توفيرها من المكتبات المساهمة، والتي تم إضافتها إلى قاعدة البيانات بين عامي (1971

---

[1] - Schoenung, J. (1981). The quality of the member-input monographic records in the OCLC on-line union catalog. Ann Arbor: University Microfilms International.

– 1977) وفحصها مقابل معيار محدد سلفًا للفهرسة يلقى قبولًا واسعًا.

تمت مقارنة التسجيلات الببليوجرافية المدخلة من قبل المكتبات الأعضاء بمثيلاتها من تسجيلات الفهرسة التي تنتجها مكتبة الكونجرس، وقد تم تحديد الانحرافات بين التسجيلة الببليوجرافية عينة الدراسة والنسخة المقابلة لها بمكتبة الكونجرس (LC) "كأخطاء"، كما تم احتساب الأخطاء بالتسجيلات الببليوجرافية في حالة عدم اتباع معايير الإدخال المقررة من قبل الـ (OCLC).

كان جزء كبير من الاختلافات بين تسجيلات المكتبات الأعضاء وتسجيلات مكتبة الكونجرس (LC) ليست أخطاء بالمعنى الحقيقي بقدر ما هي ناتجة عن التغيرات في معايير الإدخال المقررة من قبل الـ (OCLC) على مر السنين.

تم استخدام أداة متعددة الأوجه تسمى مخطط فئات الأخطاء (ECS - The Error Categorization Scheme) لتصنيف الأخطاء وفقا للموقع، والنوع، والسبب، والنتيجة.

تم ترميز الأخطاء والاختلافات بين تسجيلات المكتبات الأعضاء ومثيلاتها بنسخة (LC) في مثل هذه الطريقة لتكون مجمعة في واحدة من أربع فئات تتعلق بتأثير الخطأ:

أولًا: تلك التي تؤثر على نقاط الوصول (قاتلة)، ثانيًا: تلك التي تنطوي على تسجيلات مكررة، مما يؤثر على وصف عمل أو تحديد الهوية على مستوى الطبعة، أو الفشل في جمع أعمال مؤلف أو الطبعات للعمل (خطيرة).

ثالثًا: أخطاء ليست قاتلة ولا خطيرة وإنما (طفيفة).

رابعًا: اختلافات أو تنوعات بين نسخة مكتبة الكونجرس (LC)، والتي لا يمكن وصفها كأخطاء، إما بسبب طبيعتها (تافهة) أو لأن المكتبة المساهمة تتمسك بمعايير الإدخال التي أُقِرَّت من الـ (OCLC) (أخرى).

وقد تم تحديد (16186) خطأ واختلاف بين نسخة تسجيلات مكتبة الكونجرس (LC) وبين التسجيلات عينة الدراسة البالغة (1107) تسجيلة، وكان متوسط الأخطاء في تسجيلات المكتبات الأعضاء (14.6) خطأ، إلا أنَّ تسعة من التسجيلات بالعينة كانت بدون أية أخطاء. وكان متوسط الأخطاء القاتلة بالتسجيلة (0.62 و0.32) أخطاء جسيمة (10.94) أخطاء طفيفة منها (7.11) في الحقول ثابتة الطول و (3.82) في الحقول متغيرة الطول و (2.74) أخطاء أخرى. تم تقسيم الأخطاء بين الحقول الثابتة بلغت (7875) والحقول المتغيرة بلغت (8311). وقد بلغت نسبة التسجيلات التي تم فهرستها في قالب خطأ حوالي (2.6%).

والقصور الأكثر وضوحًا بقاعدة بيانات (OCLC) هو ارتفاع معدل التسجيلات المكررة، وتشير التقديرات إلى أن ما يصل إلى واحدة من ستة تسجيلات هي تسجيلة مكررة، والنتائج المتعلقة بالإتاحة تشير إلى أنه باستثناء البحث بالموضوع الذي لا يتوفر في الـ (OCLC)، والوصول إلى

قاعدة البيانات في معظم الوسائل متفوقة على تلك المتاحة في الفهارس البطاقية التقليدية، وجميع المتغيرات التي تم قياسها مرتبطة بشدة بجودة التسجيلات الببليوجرافية.

**Miksa, S. D. (2011)** [1]

تعمل هذه الدراسة على اكتشاف مفهوم الجودة في الفهرسة في المكتبات، وتقوم أيضا بدراسة مفهوم جودة الفهرسة بين المفهرسين الذين يمارسون عملهم في المكتبات الأكاديمية، وقد كشف البحث عن مفهوم جودة الفهرسة في أدب علوم المكتبات، أنه على الرغم من وجود شبه موافقة بالإجماع على تعريف هذا المفهوم إلا أن درجة تفصيل وتركيز هذه التعريفات تتنوع عادةً.

تم فحص هذه المفاهيم المتنوعة بدقة من أجل تطوير نظام لتقييم تعريفات جودة الفهرسة، وقد تم استخدام هذا النظام في تقييم ودراسة تعريفات المشاركين لمفهوم جودة الفهرسة، عن طريق دراسة مفاهيم المفهرسين عن جودة الفهرسة؛ لأن المفهرسين (الأصليين على وجه الخصوص) هم المسؤولون بشكل كبير عما تشتمل عليه التسجيلات الببلوجرافية.

قدم المشاركون في الاستطلاع وكان عددهم (N=296) تعريفهم الشخصي لجودة الفهرسة، وكذلك وجهات نظرهم وآراءهم في الفهرسة بأقسامهم، ومدى تأثيرها على الخطط والإجراءات التي يتبعها القسم، والبيانات المحددة التي يجب أن تشتمل عليها التسجيلة الببليوجرافية، أما المشاركون في المقابلة فقد أعطوا فكرة عن كيفية تكوين آرائهم بخصوص جودة الفهرسة والتأثيرات التي شكّلت هذه الآراء.

**American Library Association. (1999)** [2]

هي أول دراسة على نطاق واسع عن فهم رؤوس الموضوعات، وقد أجريت الدراسة بتوصية من مؤتمر عقد بمكتبة الكونجرس (LC) عن تفريعات رؤوس الموضوعات والذي اقترح توحيد نظام التفريعات لغرض تبسيط الفهرسة الموضوعية.

وكان الهدف من الدراسة تحديد إلى أي مدى يفهم مستخدمي المكتبة والمكتبيين تفريعات رؤوس الموضوعات، وأثر ذلك الفهم على استخدامهم لرؤوس الموضوعات، كما هدفت الدراسة إلى تحديد طرق أفضل لتفريع رؤوس الموضوعات، مع تحديد معايير ثابتة لبناء الرؤوس الفرعية واقترحت الدراسة بعض الطرق لتحسين فهم رؤوس الموضوعات.

خلصت الدراسة إلى أن وجود معايير محددة ومتفق عليها لبناء تفريعات رؤوس الموضوعات يؤثر على فهم واستيعاب رؤوس الموضوعات من قبل المستفيدين وأمناء المكتبات،

---

[1] - Miksa, S. D. (2011) A study of the perception of cataloging quality among catalogers in academic libraries (Doctoral dissertation, University of North Texas).

[2] - American Library Association. (1999) Understanding subject heading in library catalogs: report on the extensive study underway to determine if users & librarians understand the subject headings in catalogs. - Chicago : ALA,

كما أن الاتفاق على معايير محددة لبناء تفريعات رؤوس الموضوعات له بالغ الأثر في توفير الجهد والوقت الذي يبذل في مراجعة رؤوس الموضوعات أو تدريب القائمين على العمليات الفنية، كما اتضح من خلال الدراسة أن اختصاصي المراجع يفهمون رؤوس الموضوعات أكثر من موظفي العمليات الفنية، ومن بين النتائج أيضًا أن هناك اختلاف بين رؤوس الموضوعات التي يستخدمها الخبراء والمتخصصون في البحث عن أوعية المعلومات، وبين تلك التي يستخدمها مسؤولي العمليات الفنية بالمكتبات؛ مما يؤثر سلبًا على البحث الموضوعي عن أوعية المعلومات.

## 0/12 التعقيب على الدراسات السابقة:

ساعد استعراض الباحث للدراسات السابقة سواء العربية منها والأجنبية ذات العلاقة بموضوع الدراسة في تكوين إطارها النظري لما اشتملت عليه تلك الدراسات من مناهج وأدوات ووسائل فضلًا عما اشتملت عليه من النتائج المهمة والتوصيات القيمة التي أثرت هذه الدراسة.

وقد قصد الباحث إلى التنوع في اختيار الدراسات السابقة ذات العلاقة بموضوع الدراسة، حيث يلاحظ أن كل واحدة منها تركز على جانب مختلف عن الجانب الذي تركز عليه الدراسة الأخرى، فبعض الدراسات ركز بشكل أساس على جانب الاسترجاع الموضوعي في الفهارس الآلية أو التقليدية، والبعض الآخر ركز على جودة التسجيلات الببليوجرافية بشكل عام مع تناول نقاط الإتاحة الموضوعية بشيء من الاختصار، كما أن بعض الدراسات التي تناولت تقييم جودة التسجيلات الببليوجرافية ومن بينها نقاط الإتاحة الموضوعية قد تناولتها في إطار قطاع معين من القطاعات المعرفية البشرية، دون التعرض للقطاعات الأخرى.

بعض الدراسات التي تناولت رؤوس الموضوعات في الفهارس، تناولتها من جانب المستخدمين والمستفيدين واتجاهاتهم المختلفة نحو استخدامهم واستيعابهم لها.

تجدر الإشارة إلى أن بعض الدراسات تناولت رؤوس الموضوعات في المكتبات التقليدية غير المنضمة إلى فهرس آلي موحد تتشارك فيه جميع المكتبات في تسجيلة ببليوجرافية واحدة للوعاء الواحد.

عليه يمكن عرض أهم الملاحظات التي برزت من خلال استعراض الباحث للدراسات السابقة فيما يلي:

**أولًا:** لوحظ من خلال استعراض الدراسات السابقة تنوع المنهجيات البحثية المستخدمة لإتمام تلك الدراسات بين المنهج التجريبي والمنهج التقييمي ومنهج دراسة الحالة والمنهج المسحي الميداني.

**ثانيًا:** تنوع طرق تحديد مجتمع الدراسة بالدراسات السابقة، وإن كان معظم تلك الدراسات يعتمد على نظام العينة في الحصول على البيانات اللازمة من المجتمع الأصلي.

**ثالثًا:** لوحظ أيضًا تركيز الدراسات السابقة وميلها إلى الدراسات الكمية التي تهتم بالتحديد الدقيق

لحجم المجتمع وعينة الدراسة وهو ما تشابهت فيه تلك الدراسة مع الدراسات السابقة.

يتضح من خلال استعراض الدراسات السابقة والأطروحات التي نوقشت على المستويين العربي أو العالمي في مجال الفهارس الآلية ورؤوس الموضوعات أنها لم تتعرض لتقييم جودة نقاط الإتاحة الموضوعية المقننة في إطار التسجيلات الببليوجرافية لجميع المكتبات الأعضاء بفهرس اتحاد المكتبات الجامعية المصرية منذ نشأته في 2007 حتى الآن، وذلك وفق قواعد وأسس ونظريات صياغة وتقنين رؤوس الموضوعات.

كما لم تتعرض أي من الدراسات السابقة لتقييم إمكانيات قوائم رؤوس الموضوعات والمكانز والأدلة الإرشادية المستخدمة من وجهة نظر المفهرسين والقائمين على عملية التحليل الموضوعي ورؤوس الموضوعات بالمكتبات الأعضاء بفهرس اتحاد المكتبات الجامعية المصرية.

وتوصلت الدراسة إلى مجموعة من النتائج والتوصيات نوجزها فيما يلي:

**أولًا: النتائج:**

توصلت الدراسة إلى مجموعة من النتائج تم توزيعها وفق مجموعة من المحاور وسيتم استعراضها كما يلي:

أولًا: التجارب العالمية للفهارس الموحدة:

1. توجد العديد من التجارب العالمية الناجحة في بناء الفهارس الآلية الموحدة إلا أن أبرز هذه التجارب وأهمها وأشهرها على الإطلاق تجربة الفهرس العالمي World Cat.
2. تعد تجربة الفهرس العربي الموحد نموذج بارز ومعبر عن الفهارس العربية الموحدة.
3. صاحب بناء الفهارس الموحدة على مستوى العالم اهتمامًا بنقاط الإتاحة الموضوعية المقننة بالتسجيلات الببليوجرافية من خلال الاهتمام بعملية الضبط الاستنادي وبناء الملفات الاستنادية وتوفير أدوات التحليل الموضوعي والتي تخضع للتطوير والتحديث المستمر مثل قائمة رؤوس موضوعات مكتبة الكونجرس.

ثانيًا: خصائص نقاط الإتاحة الموضوعية المقننة بالفهارس الآلية:

4. التسجيلات الببليوجرافية في مارك تحتوي على ثلاثة أنواع من حقول الإتاحة الموضوعية هي: (XX6) التي تحتوي على المصطلحات الموضوعية المعيارية، والحقول من (050) إلى (088) وتحتوي على أرقام التصنيف وأرقام الاستدعاء ومعظم حقول الوصف من (XX2) إلى (XX5) توفر إتاحة محتملة للموضوعات باستخدام البحث بالكلمات المفتاحية.
5. تبين أيضًا من خلال دراسة خصائص نقاط الإتاحة الموضوعية وفق شكل الاتصال مارك حدوث تغييرات وصدور تحديثات بين فترة وأخرى تتوفر عليها الجهات المتخصصة بمكتبة الكونجرس الأمريكية والمسؤولة عن التحديثات المتعلقة بشكل

الاتصال مارك وخاصة فيما يتعلق بنقاط الإتاحة الموضوعية المقننة والتي كان آخر التحديثات التي جرت عليها في 2014.

ثالثًا: مشروع المكتبة الرقمية: الفهرس الموحد لاتحاد المكتبات الجامعية المصرية:

6. يتوفر على إدارة مشروع الفهرس الموحد لاتحاد المكتبات الجامعية المصرية نخبة من الأكاديميين ذوي التخصص في المكتبات والمعلومات مع مجموعة من الموظفين عددهم أربعة يمثلون أعضاء فريق ضبط الجودة بالمشروع.

7. هناك متابعة مستمرة للتحديثات التي تطرأ على شكل الاتصال مارك، من قبل وحدة المكتبة الرقمية؛ ولكن دون وضوح آلية المتابعة وفتراتها ومن الأفراد المختصين بمتابعة تلك التحديثات؟

8. ارتفاع نسب وأعداد غير المتخصصين بالمكتبات المساهمة بمشروع الفهرس كما يفتقر العاملون بالمكتبات الأعضاء بمشروع الفهرس إلى عنصر الخبرة الممتدة.

رابعًا: التحليل الموضوعي ورقابة جودة نقاط الإتاحة الموضوعية المقننة:

9. قائمة رؤوس الموضوعات الكبرى للدكتور شعبان عبد العزيز خليفة هي القائمة المعتمدة بمشروع فهرس اتحاد المكتبات الجامعية المصرية للتحليل الموضوعي للكتب العربية، وقائمة رؤوس موضوعات مكتبة الكونجرس (LCSH) للتحليل الموضوعي لأوعية المعلومات الأجنبية.

10. لا يتوفر ملف استناد آلي للموضوعات بفهرس اتحاد المكتبات الجامعية المصرية كما أنه ليس هناك نية لإنشائه على المدى القريب. وليس هناك أي وجه من أوجه التعاون بين فهرس اتحاد المكتبات الجامعية المصرية وبين المرافق الببليوجرافية الأخرى للاستفادة منها في خلق وتطوير ملفات استنادية آلية تخدم رؤوس الموضوعات العربية.

11. تعد الأدلة الإرشادية جزءًا من أدوات التحليل الموضوعي المستخدمة بفهرس اتحاد المكتبات الجامعية المصرية وقد صدر منها ست طبعات منذ بدء المشروع في ديسمبر (2007) وحتى ديسمبر (2012).

12. يشكو العاملين بالمكتبات من عدم كفاية الدورات التدريبية الممنوحة لهم وخاصة فيما يتعلق، بضيق وقت الدورات التدريبية وعدم إعطاء الأمثلة الكافية للتدريب.

13. توفر الحماسة الكافية لدى المفهرسين بالمكتبات الأعضاء بمشروع الفهرس في تطوير العمل بإجراءات التحليل الموضوعي.

14. لعدم كفاءة وكفاية الأدوات المتاحة للتحليل الموضوعي بمشروع الفهرس يلجأ المفهرسون إلى مصادر أخرى للحصول على رؤوس الموضوعات اللازمة والمعبرة عن مضامين أوعية المعلومات مما أثر سلبًا على جودة نقاط الإتاحة الموضوعية وتعدد الصيغ

المستخدمة لرأس الموضوع الواحد مما أثر على كفاءة الاسترجاع الموضوعي بالفهرس.

خامسًا: أداء فهرس اتحاد المكتبات الجامعية المصرية من الناحية الموضوعية:

15. هناك ضعف أداء فهرس اتحاد المكتبات الجمعية المصرية من الناحية الموضوعية من بين مؤشراته ما يلي:

1/15. انخفاض معدل أعداد رؤوس الموضوعات التي يتم توفيرها للكتاب الواحد بفهرس اتحاد المكتبات الجامعية المصرية.

2/15. ارتفاع أعداد التسجيلات الببليوجرافية التي تحمل رأس موضوع واحد بشكل عام بفهرس اتحاد المكتبات الجامعية المصرية.

3/15. ارتفاع نسب وأعداد التسجيلات الببليوجرافية التي تحتاج إلى رؤوس موضوعات إضافية.

4/15. ارتفاع كبير في أعداد رؤوس الموضوعات المستخدمة لمرة واحدة ونقص شديد في نسب وأعداد رؤوس الموضوعات المستخدمة لأكثر من مرة. مما يعد مؤشرًا خطيرًا على ضعف أداء فهرس اتحاد المكتبات الجامعية المصرية وقدرته على الاسترجاع الموضوعي.

5/15. ارتفاع نسب وأعداد رؤوس الموضوعات غير المطابقة للمحتوى الفعلي لأوعية المعلومات.

16. ترجع أسباب ضعف أداء فهرس اتحاد المكتبات الجامعية المصرية من الناحية الموضوعية إلى عدم الدقة وضعف الخبرة لدى المفهرسين في اختيار أكثر رؤوس الموضوعات تعبيرًا عن المحتوى الموضوعي للوعاء.

سادسًا: مدى الالتزام بأسس ومبادئ اختيار رؤوس الموضوعات بفهرس اتحاد المكتبات الجامعية المصرية:

17. هناك إخلال بمبادئ اختيار رؤوس الموضوعات بفهرس اتحاد المكتبات الجامعية المصرية تمثلت فيما يلي:

1/17. تعدد الصيغ لرأس الموضوع الواحد بفهرس اتحاد المكتبات الجامعية المصرية مما يؤدي إلى التعبير عن نفس المدلول بألفاظ وصيغ مختلفة.

2/17. العديد من الأخطاء اللغوية والإملائية في نقاط الإتاحة الموضوعية والتي لا مبرر لها سوى الإهمال من قبل القائمين على الفهرسة والتحليل الموضوعي بالمكتبات الأعضاء بمشروع الفهرس.

3/17. ارتفاع أعداد ونسب رؤوس الموضوعات المكونة من كلمة واحدة. وتعد قطاعات (العلوم التطبيقية– الفنون– العلوم الاجتماعية – الرياضيات والعلوم البحتة – الجغرافيا والتاريخ والتراجم – الديانات) الأعلى نسبة في الاحتواء على رؤوس موضوعات مكونة من كلمة واحدة.

4/17. افتقار رؤوس الموضوعات بفهرس اتحاد المكتبات الجامعية المصرية إلى درجة أو أخرى

من درجات التخصيص.

5/17. ارتفاع أعداد ونسب رؤوس الموضوعات التي تحتاج إلى درجة أو أكثر من درجات التخصيص في قطاعات (العلوم التطبيقية- العلوم الاجتماعية – الرياضيات والعلوم البحتة - الجغرافيا والتاريخ والتراجم – الديانات)

6/17. ارتفاع أعداد ونسب الرؤوس غير المقننة بفهرس اتحاد المكتبات الجامعية المصرية حيث بلغ عدد الرؤوس غير المقننة (621) رأس غير مقنن بنسبة (32.9%).

18. الانخفاض الواضح في نسب وأعداد رؤوس الموضوعات (المعقدة والمقلوبة وبين أقواس) والتي بلغت (7.9%) وهي تعد ميزة تحسب لفهرس اتحاد المكتبات الجامعية المصرية.

سابعًا: خصائص حقول الإتاحة الموضوعية المقننة بفهرس اتحاد المكتبات الجامعية المصرية:

19. يحتل الحقل (650 رأس موضوع مصطلح موضوعي) أعلى نسبة تكرار مقارنة بحقول الإتاحة الموضوعية المقننة الأخرى

20. انخفاض واضح في أعداد ونسب حقول الإتاحة الموضوعية المقننة (600 – 630 – 651) وغياب أخرى (610- 611) مما يؤثر بالسلب على كفاءة الاسترجاع بفهرس اتحاد المكتبات الجامعية المصرية

21. توجد بعض الأخطاء بالأدلة الإرشادية ومعايير ضبط الجودة الصادرة عن وحدة المكتبة الرقمية بالمجلس الأعلى للجامعات.

ثانيًا: التوصيات:

استنادًا إلى النتائج السابقة توصي الدراسة بما يلي:

1- الأخذ بنتائج هذه الدراسة لتحسين جودة نقاط الإتاحة الموضوعية المقننة والارتقاء بكفاءة الاسترجاع الموضوعي بفهرس اتحاد المكتبات الجامعية المصرية.

2- الإسراع في تعديل الوضع الوظيفي سواء للعاملين بفريق ضبط الجودة بوحدة المكتبة الرقمية أو العاملين بالمكتبات الأعضاء من النظام المؤقت إلى النظام الدائم للحفاظ على الأفراد ذوي الخبرات الممتدة والكفاءات.

4- السعي بشكل أكبر في توظيف ذوي الخبرة والتخصص في مجال المكتبات والمعلومات والتقليل قدر المستطاع من الاستعانة بغير المتخصصين وخاصة في القائمين بإجراءات التحليل الموضوعي.

5- تكثيف الدورات التدريبية في مجال التحليل الموضوعي لرفع مستوى وأداء العاملين بالفهرسة والتحليل الموضوعي والأخذ بعين الاعتبار توفير الوقت الكافي للدورات التدريبية ومنح فرصة أكبر لمزيد من الأمثلة التوضيحية التي تعرف أكثر بأسس ومبادئ صياغة رؤوس الموضوعات.

6- استكمال توفير أدوات التحليل الموضوعي في جميع المكتبات الأعضاء على المدى القريب

والعمل على توفير هذه الأدوات في الشكل الإلكتروني والاستعانة، بالموردين تعد من أفضل السبل لتحقيق هذا الهدف المهم والحيوي.

7- الإسراع في بناء ملف استناد آلي للموضوعات لفهرس اتحاد المكتبات الجامعية المصرية بوضع تصور واضح للخطوات التي ينبغي أن تتم في هذا الإطار ويتوفر عدد من الطرق التي يمكن الاختيار من بينها أو المفاضلة فيما بينها بما يناسب إمكانات مشروع فهرس اتحاد المكتبات الجامعية المصرية وهي

1/7 الاعتماد على جهود العاملين بأقسام العمليات الفنية والمفهرسين بالمكتبات الاعضاء بمشروع الفهرس لإنشاء تسجيلات استناديه لتكون نواة لبناء ملف استناد آلي للموضوعات.

2/7 أو الاستعانة بموردي النظم الآلية والذين يقدمون خدمات بناء وصيانة ملف استنادية للموضوعات (وهو الاتجاه الذي تسير فيه أغلب التجارب العالمية لبناء وصيانة الملفات الاستنادية).

8- إنشاء روابط تعاون حقيقية مع الجهات المناظرة والمرافق الببليوجرافية العربية في مجال الضبط الاستنادي الآلي.

9- التأكيد على المفهرسين في توفير عدد نقاط إتاحة موضوعية أكثر لما لوحظ من ارتفاع معدلات التسجيلات الببليوجرافية التي تحتوي على رأس موضوع واحد.

10- اعتماد آلية مختلفة من قبل فريق ضبط الجودة بوحدة المكتبة الرقمية لمراقبة جودة نقاط الإتاحة الموضوعية المقننة بفهرس اتحاد المكتبات الجامعية المصرية.

11 – مراقبة الطريقة التي تنمو بها نقاط الإتاحة الموضوعية المقننة بفهرس اتحاد المكتبات الجامعية المصرية والحد من ظاهرة النمو غير المبرمج أو المنظم والزيادة غير المبررة أحيانًا في عدد رؤوس الموضوعات – المستخدمة في التسجيلة الببليوجرافية الواحدة.

12- الاستفادة من الحماسة العالية والمتوفرة لدى المفهرسين لتحسن جودة نقاط الإتاحة الموضوعية والرغبة في وجود ملف استناد آلي للموضوعات يتم الاستعانة به في اجراءات التحليل الموضوعي بالفهرس.

13- التواصل مع أقسام المكتبات بالجامعات المصرية والتعاون معها لتحقيق ما يلي :

1/13 الاستعانة بالأساتذة والمتخصصين والطلاب في كل جامعة للتعاون مع فريق ضبط الجودة بها لمراجعة نقاط الإتاحة الموضوعية وتقديم المساعدة اللازمة لتحسينها.

2/13 مطالبة الأساتذة بالتركيز على تطوير المناهج المقدمة في مجال التحليل الموضوعي وربطها بجزء عملي يتعلق بالتعامل مع نقاط الإتاحة الموضوعية بفهرس اتحاد المكتبات الجامعية المصرية.

14- إعداد دليل إرشادي مفصل يختص بنقاط الإتاحة الموضوعية المقننة على أن يأخذ في

الحسبان النتائج والتوصيات التي توصلت إليها الدراسة للفت انتباه المفهرسين نحو النقاط المهمة التي ينبغي التركيز عليها.

15- ضرورة الفحص الفعلي الكافي لأوعية المعلومات من قبل المفهرسين لاستخلاص الجوانب الموضوعية المهمة دون الاعتماد على عناوين الأوعية أو الاعتماد على بطاقة الفهرسة أثناء النشر والتي لا يعرف مصدرها أو الطريقة التي أعدت بها.

16- ضرورة الرجوع إلى القائمة المعتمدة في حالة وجودها لتجنب ارتفاع نسب وأعداد رؤوس الموضوعات غير المقننة بفهرس اتحاد المكتبات الجامعية المصرية.